\documentclass{desyproc}

\usepackage[outercaption]{sidecap}

\begin{document}
\title{A First Look at the Impact of NNNLO Theory Uncertainties on Top Mass Measurements at the ILC}

\author{{\slshape Frank Simon\footnote{email: fsimon@mpp.mpg.de\\Talk presented at the International Workshop on Future Linear Colliders (LCWS15), Whistler, Canada, 2-6 November 2015}}\\[1ex]
Max-Planck-Institut f\"ur Physik, F\"ohringer Ring 6, 80805 M\"unchen, Germany}

\acronym{LCWS 2015} 
\maketitle


\begin{abstract}
A scan of the top production threshold at a future electron-positron collider provides the possibility for a precise measurement of the top quark mass in theoretically well-defined mass schemes. With statistical uncertainties of 20 MeV or below, systematics will likely dominate the total uncertainty of the measurement. This contribution presents a first look at the impact of the renormalization scale uncertainties in recent NNNLO calculations of the top pair production cross section in the threshold region on the measurement of the top quark mass at the International Linear Collider. 
\end{abstract}

\section{Introduction}

Due to its high mass, the top quark has only been studied at hadron colliders up to now. The most precise direct measurements of its mass today come from the LHC, with total experimental uncertainties of as low as 500 MeV \cite{Khachatryan:2015hba},  dominated by systematic uncertainties. Since these measurements provide the top quark mass in the context of the event generator used in the analysis, additional uncertainties, currently estimated to be on the order of 1 GeV, are incurred when transforming the measured mass value to theoretically well-defined mass schemes as used in precision calculations. 

A scan of the top pair production cross section at an electron-positron collider holds the potential for a precise measurement of the mass of the top quark in a theoretically well-defined framework.  An experimental study \cite{Seidel:2013sqa} in the context of the future linear collider projects ILC \cite{Behnke:2013xla} and CLIC \cite{Lebrun:2012hj} based on detailed detector simulations has shown that statistical uncertainties of 20 MeV on the top quark mass are reached with an integrated luminosity of 100 fb$^{-1}$. In typical running scenarios for these projects, this or higher integrated luminosities are foreseen for a threshold scan. At this level of precision, experimental and theoretical systematics are highly relevant. In \cite{Seidel:2013sqa} and \cite{Simon:2014hna} experimental systematics are estimated to be on the level of 30 - 50 MeV. The precision of the strong coupling constant also enters in the analysis, with a corresponding uncertainty of 16 MeV based on the 2014 uncertainty of the world average of $\alpha_s$ \cite{Agashe:2014kda}. In \cite{Seidel:2013sqa}, a naive estimate of the theory uncertainty, assuming an overall scale uncertainty of the calculation of the cross section of 3\% as used also in \cite{Martinez:2002st}, has yielded an uncertainty of the top mass of 56 MeV. This suggests that theory uncertainties will be among the leading uncertainties, making a further study of this issue of high importance. 

The recent completion of NNNLO QCD calculations \cite{Beneke:2015kwa} of the top pair production cross section, including NLO non-resonant electroweak contributions \cite{Beneke:2010mp, Penin:2011gg}, NNNLO Higgs effects and QED \cite{Beneke:2015lwa, Beneke:2013kia}, enables the study of theory uncertainties for calculations with a precision which is unlikely to be substantially improved by the time such experimental measurements are performed at a future $e^+e^-$ collider. Here, the results of these calculations are combined with the experimental techniques developed in \cite{Seidel:2013sqa} to perform a first preliminary analysis of the impact of the scale uncertainties on the top mass extraction.

\section{The impact of scale uncertainties on the cross section}

\begin{figure}[ht]
\centering
\includegraphics[width=0.495\textwidth]{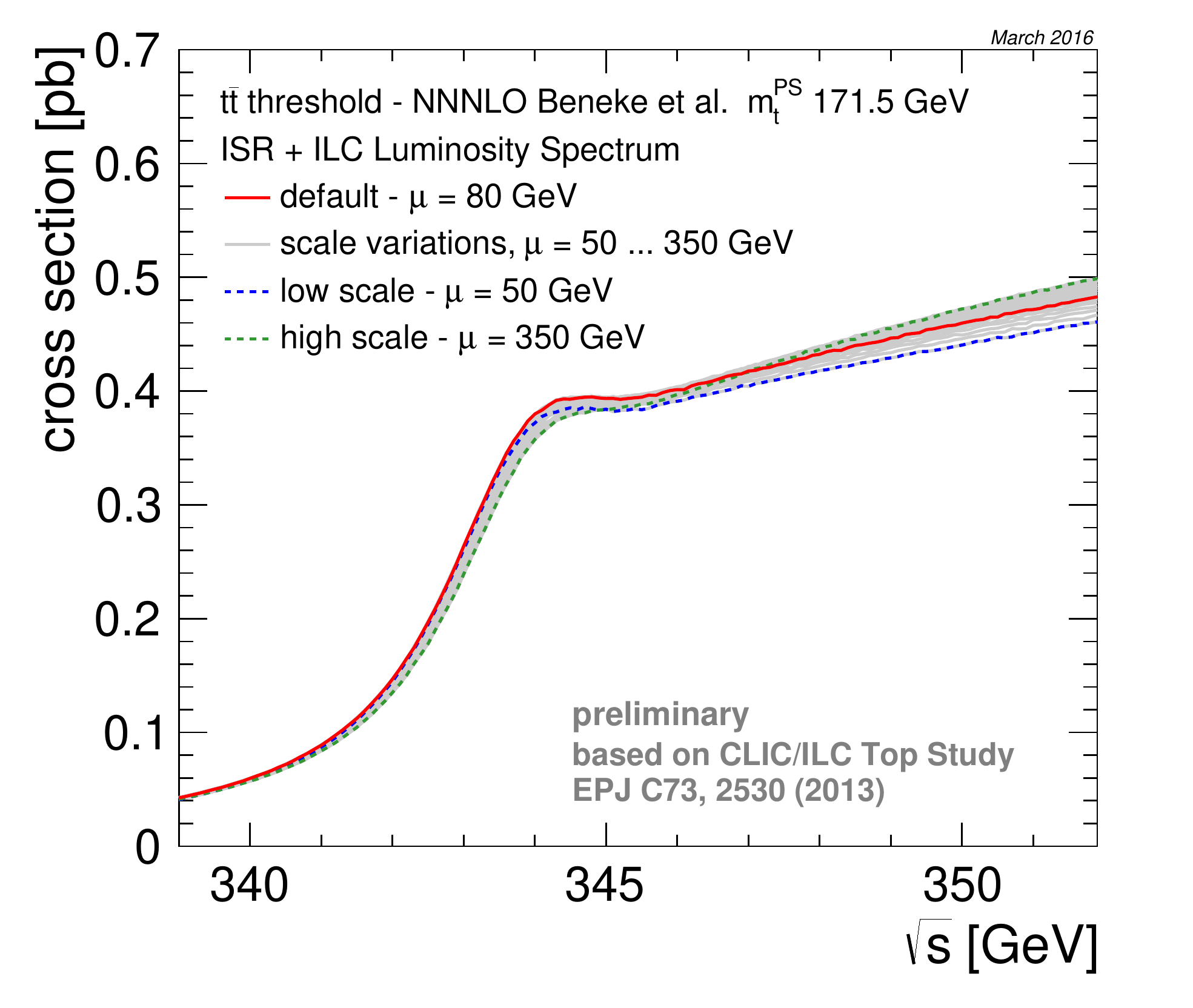}
\hfill
\includegraphics[width=0.495\textwidth]{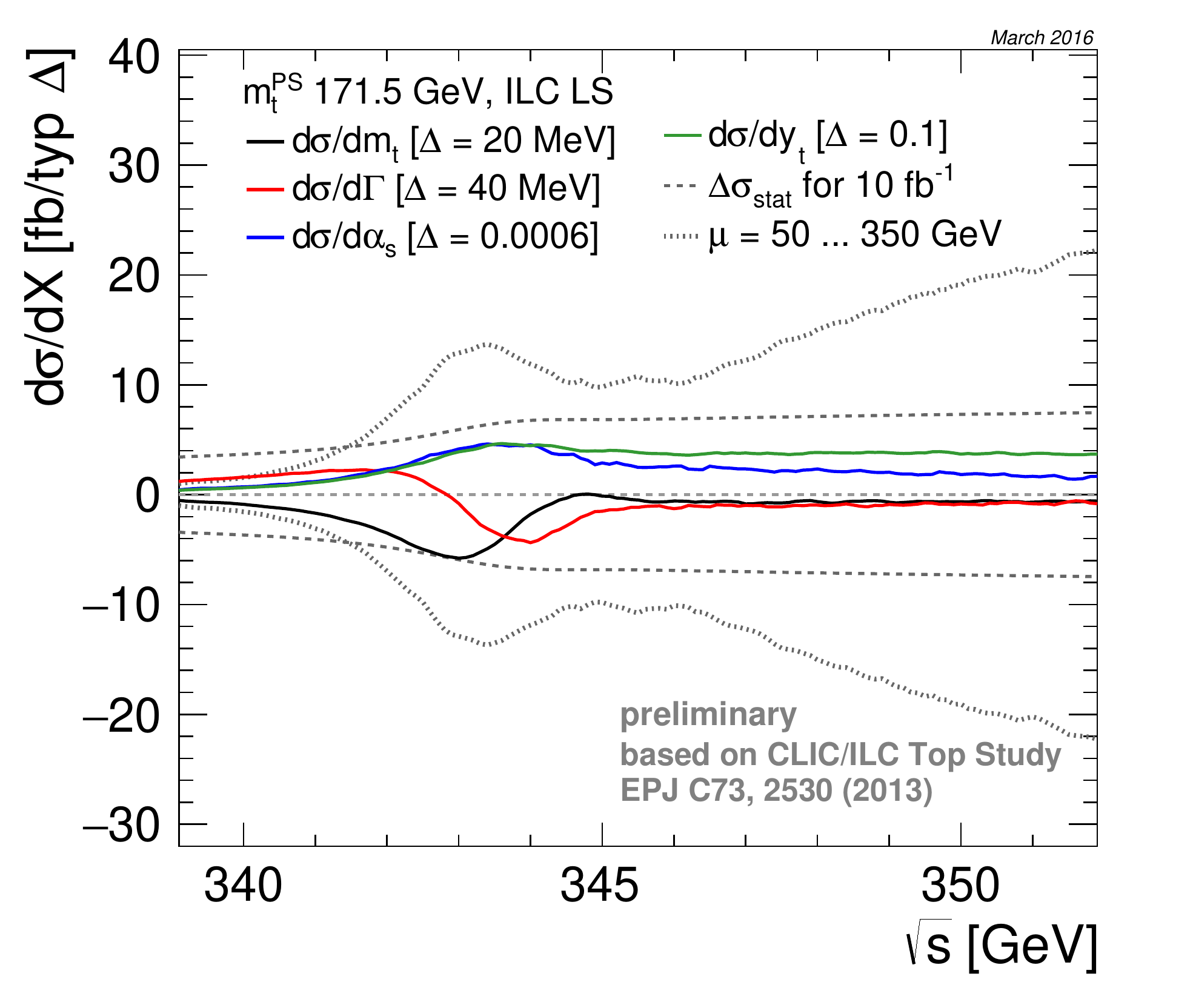}
\caption{{\it Left:} Top pair production cross section with NNNLO precision \cite{Beneke:2015kwa, Beneke:2015lwa}, including ISR and ILC luminosity spectrum effects, also showing scale variations over the range considered in the present analysis. {\it Right:} Variation of the cross section for changes of different parameters, in the context of the scale uncertainties and the statistical precision of a single \mbox{10 fb$^{-1}$} data point. The range of changes of the different parameters is given by typical projected statistical uncertainties for the different quantities as listed in the figure.}\label{fig:TopXsSystematics}
\end{figure}

For the present study, a top quark mass of $m_t^{\mathrm{PS}}$ = 171.5 GeV in the potential-subtracted mass \mbox{scheme \cite{Beneke:1998rk}} and a top quark width of 1.33 GeV is assumed. Figure \ref{fig:TopXsSystematics} {\it left} shows the calculated cross section as a function of collider center of mass energy, taking into account  corrections due to initial state radiation (ISR) and due to the luminosity spectrum of the ILC. Also shown are the variations of the cross section induced by varying the renormalization scale $\mu$ in the range of 50 GeV to 350 GeV, as discussed in detail in \cite{Beneke:2015kwa}. 

Figure \ref{fig:TopXsSystematics} {\it right} illustrates the effect of variations of the top quark mass $m_t$, the top quark width $\Gamma_t$, the strong coupling constant $\alpha_s$ and the top Yukawa coupling $y_t$ on the production cross section. To allow a simple visual interpretation of these variations, the are shown for typical statistical uncertainties projected for $e^+e^-$ collider measurements of the various quantities, as listed in detail in the figure. Note that the variation of $\alpha_s$ corresponds to the uncertainty of the world average quoted in 2014 \cite{Agashe:2014kda}, prior to the increase of the uncertainty in the latest evaluation \cite{BethkeAlphas}. Also shown are the statistical uncertainties of a single 10 fb$^{-1}$ data point and the size of the scale variations. The latter are symmetrized to provide better visibility. It is apparent that the mass can be extracted in a region that is less affected by changes in other properties, while in particular the effects of the top Yukawa coupling and the strong coupling are highly correlated, resulting in difficulties of extracting the former without substantial improvements in the precision of the latter. It is also apparent that the scale variations are larger than the variations induced by changes of top quark parameters within the range of the projected  uncertainties, suggesting theory systematics in excess of the statistical uncertainties.

\section{A threshold scan analysis with NNNLO scale uncertainties}

To provide a first look at the impact of the NNNLO scale uncertainties on the measurement of the top quark mass in a threshold scan, the analysis discussed in detail in \cite{Seidel:2013sqa} has been extended, as discussed in the following. The precision with which the mass can be measured is determined by a toy MC study which simulates a large number of threshold scans. For each iteration, data points are generated at the energy points considered in the study, using reconstruction efficiencies and background levels determined from full simulation studies combined with the input cross section. From the simulated measurement points the top quark mass is determined by a template fit, comparing the data points to calculated cross section curves for different mass hypotheses. In this study, the top quark width and the Yukawa coupling are assumed to take their Standard Model values and are not varied. The strong coupling constant is assumed to be an external input. The impact of the uncertainty of $\alpha_s$ is briefly discussed in Section \ref{sec:Alpha}.

\begin{figure}[ht]
\begin{minipage}[t]{0.5\textwidth}
 \vspace{0pt}
\includegraphics[width=0.99\textwidth]{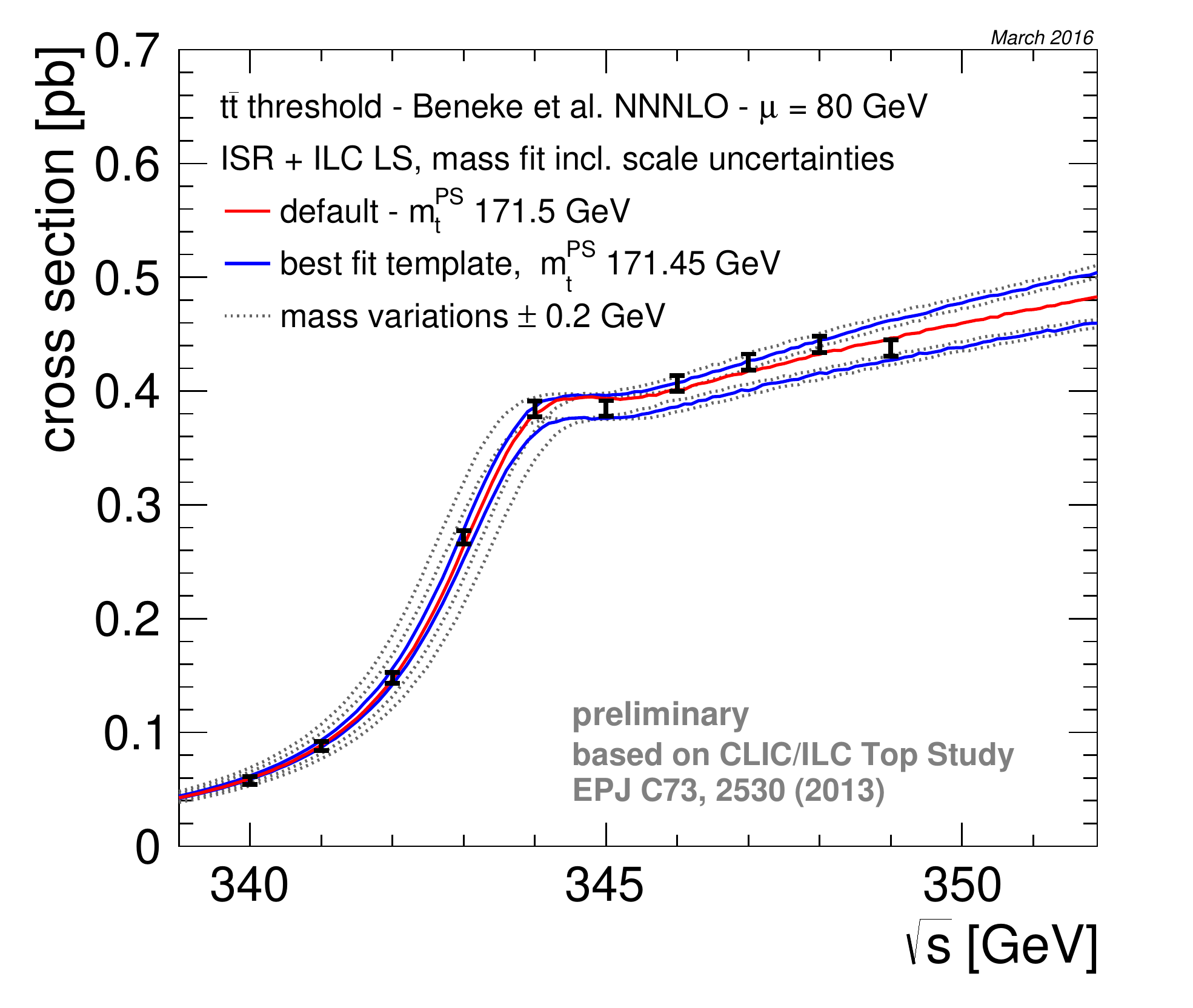}
\end{minipage}
\begin{minipage}[t]{0.49\textwidth}
 \vspace{0pt}
\caption{Illustration of a $t\bar{t}$ production threshold scan, showing the best fit template for a fit accounting for NNNLO scale variations and the variations of the template for mass changes of $\pm$ 100 MeV.}\label{fig:ThresholdScan}
\end{minipage}
\end{figure}

In the present study, the template fit is extended to account for the NNNLO scale uncertainties. Instead of comparing the simulated data points with a line given by the cross section for a particular mass hypothesis, the comparison is now made with a band that is defined by the space between the minimum and the maximum cross section over the range of considered renormalization scales at each energy point. This is illustrated in Figure \ref{fig:ThresholdScan}. The simulated data points correspond to 10 points with an integrated luminosity of 10 fb$^{-1}$ each, using the default cross section for $m_t^{\mathrm{PS}}$ = 171.5 GeV and $\mu$ = 80 GeV as input. The template providing the best fit, in this example corresponding to a top quark mass of 171.45 GeV, is given by the area between the two solid blue lines. Also shown are the templates for shifts in mass by $\pm$100 MeV, shown by the dotted lines. To account for the bands used as templates, the calculation of the $\chi^2$ used in the template fit is modified. Data points that lie within the template they are compared to do not increase the global $\chi^2$ used in the fit, for data points beyond the band the $\chi^2$ is increased by $\Delta\chi^2\, =\, \Delta\sigma^2/\Delta_{\mathrm{data}}^2$, where $\Delta\sigma$ is the difference between the measured cross section and the nearest edge of the template band, and $\Delta_{\mathrm{data}}$ is the statistical uncertainty of the data point. 

\begin{figure}[ht]
\centering
\includegraphics[width=0.495\textwidth]{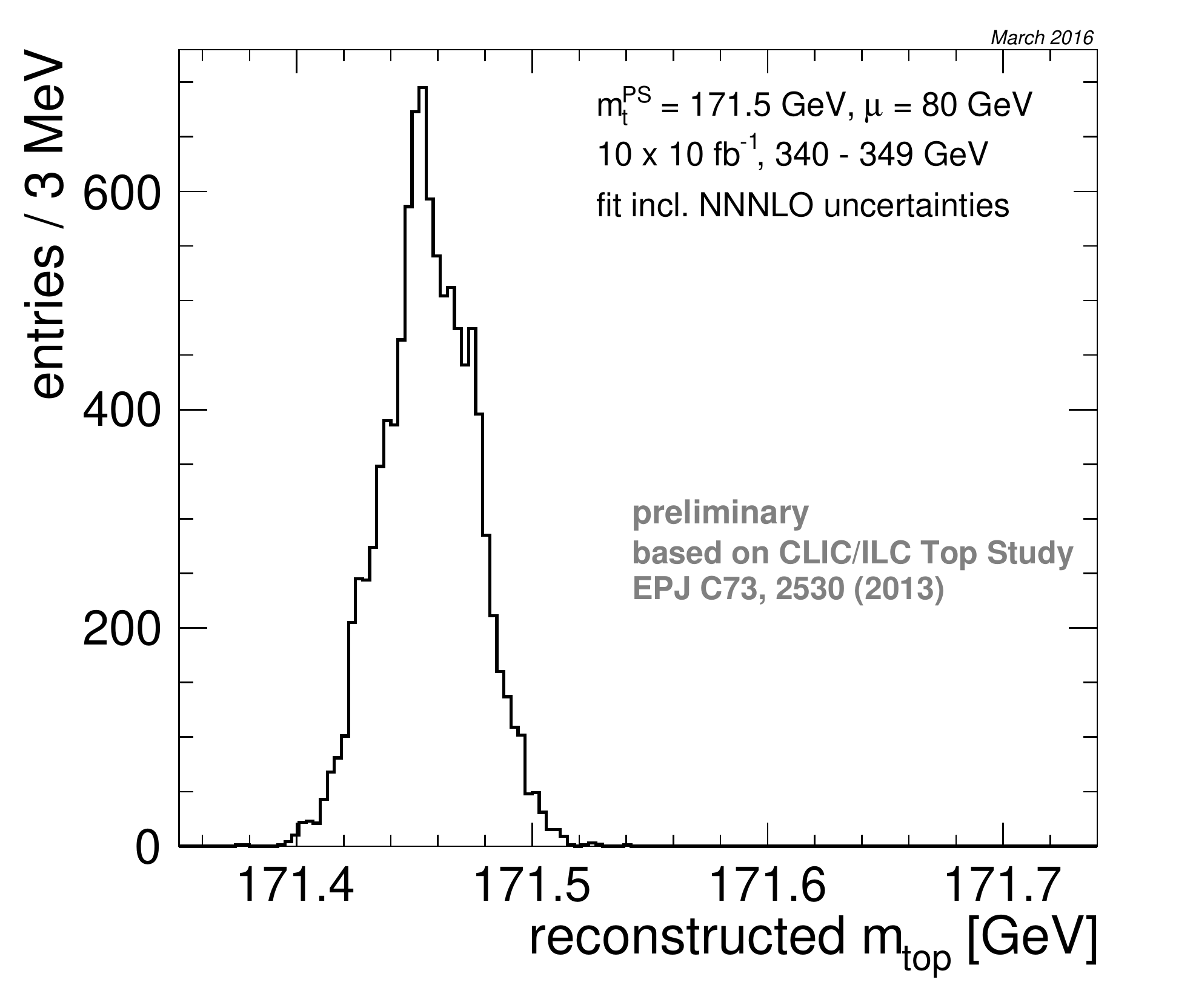}
\hfill
\includegraphics[width=0.495\textwidth]{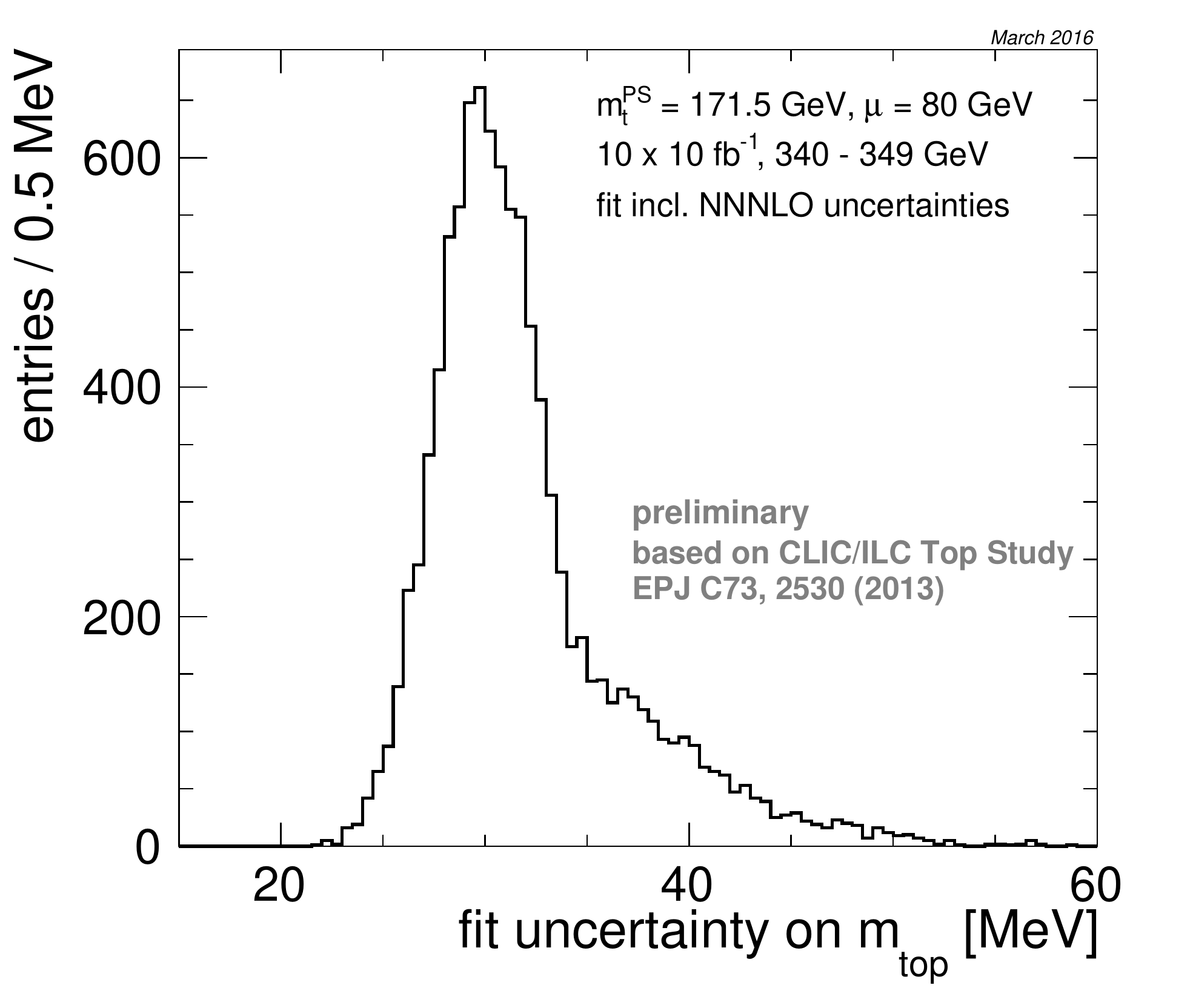}
\caption{{\it Left:} Distribution of the extracted mass in a template fit accounting for NNNLO scale variations for a ToyMC study with 10\,000 iterations for an input mass value of 171.5 GeV. The observed bias of approximately 50 MeV is due to the asymmetry of the impact of the scale variations as discussed in the text. {\it Right:}  Distribution of the fit uncertainty determined from the $\chi^2$ distribution of the template fit for 10\,000 iterations. The mean fit uncertainty is 32 MeV, with a most probable uncertainty of 30 MeV.}\label{fig:FitResults}
\end{figure}

Since the template bands are not symmetric around the cross section for the default value for the renormalisation scale of $\mu$ = 80 GeV, the fitted mass value is biased to lower top masses by 44 GeV. This bias represents the choice of the $\mu$ parameter in the fit, and is thus a trivial offset that can be corrected for. Figure \ref{fig:FitResults} {\it left} shows the fit results for 10\,000 iterations. The standard deviation of the distribution is 19 MeV, representing the statistical uncertainty of the fitted mass. This is only marginally higher than the same value for a fit that does not include the scale variations and is instead using just a single line per mass calculated for $\mu$ = 80 GeV. For a single analysed threshold scan the full uncertainty of the fit is larger, however, since not all data points contribute to the $\chi^2$. This uncertainty is determined by the variations in mass that lead to an increase of the global $\chi^2$ by 1. The numerical value is determined by a parabolic fit to the $\chi^2$ distribution of the template fit. For cases where the minimum of the parabola is below 0, the mass values which result in a $\chi^2$ of 1 are taken as the bounds of the uncertainties. 

The distribution of the full fit uncertainty for 10\,000 iterations of the toy MC is given in Figure \ref{fig:FitResults} {\it right}. In the present study, the mean fit uncertainty is 32 MeV, with a most probable uncertainty of 30 MeV, and 90\% of all iterations resulting in an uncertainty of less than 39 MeV. Assuming that the fit uncertainty is given by the quadratic sum of a purely statistical component and of an additional fit uncertainty originating from the use of the NNNLO scale uncertainties in the template fit which does not depend  on the integrated luminosity, the additional fit uncertainty is 25 MeV. Tests with different integrated luminosities have shown that while this simplified representation is not entirely accurate, it allows a good extrapolation of the expected uncertainties with an accuracy of better than 10\% for changes of integrated luminosities of a factor of two. Larger deviations are only seen for cases with very small integrated luminosities.

\section{Systematic uncertainties derived from scale variations}
\label{sec:Scale}

Based on the analysis discussed above, the theory uncertainties associated with the measurement of the top quark mass in a threshold scan are evaluated. This is done by performing the analysis described in the previous section taking different values of the renormalization scale $\mu$ for the calculation of the input cross section.

\begin{figure}[ht]
\begin{minipage}[t]{0.5\textwidth}
 \vspace{0pt}
\includegraphics[width=0.99\textwidth]{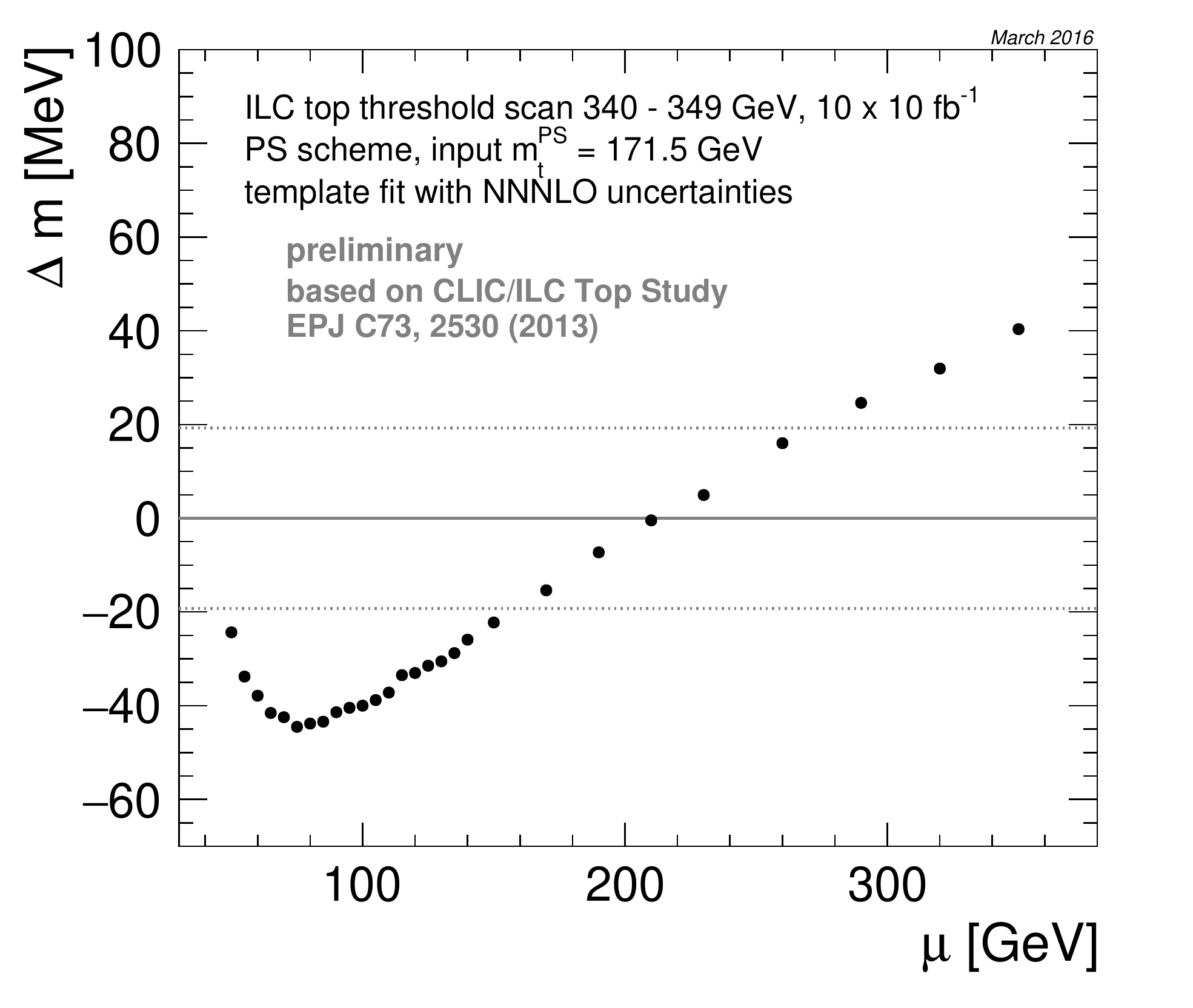}
\end{minipage}
\begin{minipage}[t]{0.49\textwidth}
 \vspace{0pt}
\caption{Distribution of the central value of the fitted mass for variations of the scale $\mu$ used to generated the input cross section. The template fit used to extract the mass includes NNNLO scale uncertainties.} \label{fig:SystematicsScan}
\end{minipage}
\end{figure}

Figure \ref{fig:SystematicsScan} shows the deviation of the mean reconstructed top quark mass as a function of the renormalization scale used for the generation of the input cross section for 10\,000 toy MC iterations per data point. As seen in the previous section, the fitted mass shows a bias of 45 MeV for  the default value of $\mu$ = 80 GeV. This is explained by the fact that the cross section for $\mu$ = 80 GeV follows the maximum of the band given by the scale variations over a wide range of center of mass energies, as shown in Figure \ref{fig:TopXsSystematics} {\it left}. Over the full range of scale variations considered here, the fitted mass varies by $\pm45$ GeV, which is taken as the theoretical uncertainty of the mass determination in a threshold scan. This uncertainty is comparable in size to the uncertainty derived for a naive 3\% scale uncertainty in the theory in \cite{Seidel:2013sqa}. This is not surprising, considering that the scale variations result in typical variations of the cross section by a similar order of magnitude, however with differences in different regions of the threshold \cite{Beneke:2015kwa}. 

\section{The impact of the strong coupling constant}
\label{sec:Alpha}

As shown in Figure \ref{fig:TopXsSystematics} {\it right}, the uncertainty of the strong coupling constant has a sizeable impact on the overall cross section, shown in the figure for the 2014 world average uncertainty of $6\,\times\,10^{-4}$.  By following the same strategy as for the NNNLO scale variations discussed in Section \ref{sec:Scale}, the impact of uncertainties of the strong coupling constant on the mass determination are studied. Here, the value of $\alpha_s$ used for the calculation of the input cross section is varied in the analysis, keeping the other parameters constant. For variations of the order of 10$^{-3}$, the bias of the reconstructed mass depends linearly on the offset in the strong coupling. The  systematic uncertainty on $m_t$ is  2.7 MeV per 10$^{-4}$ uncertainty of $\alpha_s$, corresponding to 16 MeV for the 2014 world average \cite{Agashe:2014kda}, and to 35 MeV for the new preliminary world average \cite{BethkeAlphas}.

\section{Summary}

This first preliminary study of the impact of the renormalization scale uncertainties of NNNLO QCD calculations on the measurement of the top quark mass in a threshold scan at the International Linear Collider shows that theory uncertainties are among the leading systematics of such a high-precision measurement. The scale uncertainties enter in the present analysis in two ways: 
\begin{itemize}
\item When accounting for the uncertainties in the template fit used to extract the mass from the measured cross section at several energy points, the templates to compare the data points to change from lines to bands representing the scale uncertainty at each energy point. Since data points that lie within the band given by the scale uncertainties do not contribute to the fit $\chi^2$, the uncertainty of the fitted mass depends on the measured values. The mean fit uncertainty for an integrated luminosity of 100 fb$^{-1}$ is 32 MeV. This can be divided into a statistical uncertainty of 19 MeV and an additional uncertainty originating from the inclusion of the scale uncertainties in the fit, which only weakly depends on the integrated luminosity, and amounts to 25 MeV.
\item The uncertainties also result in an overall systematic uncertainty of the measured mass, with the result of the measurement varying over a range of 90 MeV for variations of the renormalization scale $\mu$ from 50 GeV to 350 GeV. Expressed as a symmetrized uncertainty, the theoretical uncertainty originating from the scale variations is thus 45 MeV. 
\end{itemize}

When taking theory uncertainties into account, the precision of the mass fit improves only very slowly with increasing statistics beyond a total integrated luminosity of 100 fb$^{-1}$. Combining the 32 MeV fit uncertainty with theoretical systematics of 45 MeV and experimental systematics estimated to be in the few 10 MeV range, it is expected that the total uncertainties of a top quark mass measurement in a threshold scan at a future linear electron-positron collider are at or below 100 MeV. 

In upcoming studies, the first preliminary analysis presented here will be further refined to put the conclusions on a more solid footing. It will also be extended to include other relevant top quark properties, in particular the top quark width and the Yukawa coupling, and will be performed for different $e^+e^-$ collider options.  
 
 \section*{Acknowledgements}
 
The author thanks Martin Beneke and Jan Piclum for helpful discussions and the sharing of code to perform the calculations of the cross sections needed in the analysis. This work was supported by the DFG cluster of excellence `Origin and Structure of the Universe'.


\begin{footnotesize}


\end{footnotesize}


\end{document}